\documentclass[showpacs,10pt,twocolumn,prb]{revtex4-1}
\usepackage{amsmath}
\usepackage{amssymb}
\usepackage{graphics}
\usepackage{epsfig}
\usepackage{CJK}

\begin{document}

\begin{CJK*}{GBK}{}

\title{Giant increase in critical current density of K$_{x}$Fe$_{2-y}$Se$%
_{2} $ single crystals}
\author{Hechang Lei and C. Petrovic}
\affiliation{Condensed Matter Physics and Materials Science Department, Brookhaven
National Laboratory, Upton, NY 11973, USA}
\date{\today}

\begin{abstract}
The critical current density $J_{c}^{ab}$ of K$_{x}$Fe$_{2-y}$Se$_{2}$
single crystals can be enhanced by more than one order of magnitude, up to $%
\sim $ 2.1$\times $10$^{4}$ A/cm$^{2}$ by post annealing and quenching
technique. A scaling analysis reveals the universal behavior of the
normalized pinning force as a function of the reduced-field for all
temperatures, indicating the presence of a single vortex pinning mechanism.
The main pinning sources are three dimensional (3D) point-like normal cores.
The dominant vortex interaction with pinning centers is via spatial
variations in critical temperature $T_{c}$ ("$\delta T_{c}$ pinning").
\end{abstract}


\maketitle
\end{CJK*}

\section{Introduction}

Recently discovered iron-based superconductors\cite{Kamihara} induce great
interest in scientific community because of rather high $T_{c}$, proximity
to the spin-density wave state and multiband nature of electronic transport.%
\cite{Ren}$^{-}$\cite{Hunte} However, these materials also encourage
potential technical applications due to high upper critical fields $\mu
_{0}H_{c2}$ and critical current densities $J_{c}$.\cite{Hunte}$^{-}$\cite%
{Yang H}

In the family of iron-based superconductors, FeCh (Ch = S, Se, and Te,
FeCh-11 type) materials have the simplest crystal structure, nearly
isotropic high $\mu _{0}H_{c2}$ and rather high $J_{c}$,\cite{Lei HC1}$^{,}$%
\cite{Yadav} but their relatively low $T_{c}$ impedes prospects for
applications. Superconducting $T_{c}$ was raised up to about 32 K in A$_{x}$%
Fe$_{2-y}$Se$_{2}$ (A = K, Rb, Cs, and Tl, FeCh-122 type) iron selenide
superconductors with rather high $\mu _{0}H_{c2}$ ($\sim $ 56 T for $%
H\parallel c$ at 1.6 K).\cite{Guo}$^{,}$\cite{Mun} Preliminary results
indicate that the $J_{c}$ of K$_{x}$Fe$_{2-y}$Se$_{2}$ is much lower when
compared to iron arsenides or binary FeCh-11 type iron selenides.\cite%
{Karpinski}$^{,}$\cite{Yang H}$^{,}$\cite{Yadav}$^{,}$\cite{Lei HC2}$^{-}$%
\cite{Gao ZS} Post annealing and quenching treatment can induce metallic and
superconducting state in as-grown and insulating K$_{x}$Fe$_{2-y}$Se$_{2}$
crystals,\cite{Han} yet current carrying characteristics of such materials
are not known.

In this work, we report on the significant enhancement of critical current
density\ in K$_{x}$Fe$_{2-y}$Se$_{2}$ single crystals obtained via
post-annealing and quenching process. We also give detailed insight into the
vortex pinning mechanism. Main pinning sources are the 3D normal cores
whereas dominant vortex interaction with pinning centers is via spatial
variations in $T_{c}$.

\section{Experiment}

Details of crystal growth and structure characterization were reported
elsewhere.\cite{Lei HC2} The as-grown crystals were sealed into Pyrex tube
under vacuum ($\sim $ 10$^{-1}$ Pa). The samples were annealed at 400 $%
^{\circ }$C for 1h and quenched in the air as reported previously.\cite{Han}
. Crystals were claved and cut into rectangular bars. Magnetization
measurements were performed in a Quantum Design Magnetic Property
Measurement System (MPMS-XL5).

\section{Results and Discussion}

\begin{figure}[tbp]
\centerline{\includegraphics[scale=0.42]{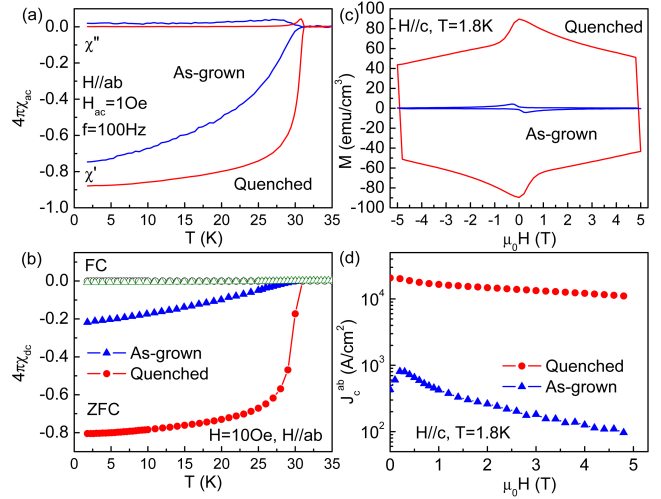}} \vspace*{-0.3cm}
\caption{Temperature dependence of the (a) ac and (b) dc magnetic
susceptibility of as-grown and quenched K$_{x}$Fe$_{2-y}$Se$_{2}$ crystals
taken in $\protect\mu _{0}H$ = 0.1 (ac) and 1 mT (dc) field, respectively.
(c) Magnetization hysteresis loops of as-grown and quenched samples at 1.8 K
for $H\parallel c$. (d) Superconducting critical current densities $%
J_{c}^{ab}(\protect\mu _{0}H)$ of as-grown and quenched samples.}
\end{figure}

Calculated volume fractions from ac susceptibility at 1.8 K are rather
similar, 75\% for as grown and 88\% for quenched crystal. However, the
quenched crystal shows a very steep transition at 31 K and saturates at
about 10 K whereas for as-grown sample the diamagnetic signal increases
gradually with slightly lower $T_{c}$ (Fig. 1(a)). The single sharp peak of 4%
$\pi \chi "$ in quenched crystals (Fig. 1(a,b)) indicates more homogeneous
superconducting state. The calculated volume fraction from dc susceptibility
(Fig. 1(b)) significantly increased after quenching, consistent with
previous results.\cite{Han} Hence, post-annealing and quenching process
singnificantly advances superconducting volume fraction in quenched K$_{x}$Fe%
$_{2-y}$Se$_{2}$. The small volume fraction estimated from the FC curve
suggests possible strong magnetic flux pinning effects.

Magnetic hyshteresis loops (MHL) of quenched sample are much bigger and more
symmetric (Fig. 1(c)). The pinning force is enhanced significantly and the
bulk pinning is dominant when compared to the as-grown sample. The MHL of
as-grown crystal is small and asymmetric, suggesting that the surface
barrier may be important.\cite{Pissas}$^{,}$\cite{Zhang L} Moreover, there
is no fishtail effect up to 5 T\ which has been observed in S-doped K$_{x}$Fe%
$_{2-y}$Se$_{2-x}$S$_{x}$ single crystal with S = 0.99 at low field and in
FeAs-122 single crystals at high field.\cite{Yang H}$^{,}$\cite{Lei HC}$^{,}$%
\cite{Sun DL}$^{-}$\cite{Prozorov}

The in-plane critical current density $J_{c}^{ab}(\mu _{0}H)$ for a
rectangularly-shaped crystal with dimension $c<a<b$ when $H\parallel c$ is%
\cite{Bean}$^{,}$\cite{Gyorgy}

\begin{equation}
J_{c}^{ab}(\mu _{0}H)=\frac{20\Delta M(\mu _{0}H)}{a(1-a/3b)}
\end{equation}

where a and b ($a<b$) are the in-plane sample size in cm, $\Delta M(\mu
_{0}H)$ is the difference between the magnetization values for increasing
and decreasing field at a particular applied field value (measured in emu/cm$%
^{3}$), and $J_{c}^{ab}(\mu _{0}H)$ is the critical current density in A/cm$%
^{2}$. As shown in Fig. 1(d), the calculated $J_{c}^{ab}(0)$ for quenched
sample from Fig. 1(c) is enhanced about 50 times when compared to as-grown
sample. This can not be simply ascribed to the improvement of the
superconducting volume fraction, because the volume fraction of quenched
crystal is only about 4 times larger than the volume fraction of the
as-grown crystal. Critical current values in quenched crystal are higher
than that in K$_{x}$Fe$_{2-y}$Se$_{2}$ crystals grown using the one-step
technique and are the highest known $J_{c}^{ab}$ among FeCh-122 type
materials.\cite{Gao ZS} The quenched sample also exhibits better performance
at high field. The $J_{c}^{ab}$ for quenched sample is still larger than 10$%
^{4}$ A/cm$^{2}$ at 4.8 T whereas for as-grown sample, it has decreased
about one order of magnitude. The $J_{c}^{ab}(4.8T,1.8K)$ is also larger
than for K$_{x}$Fe$_{2-y}$Se$_{2-z}$S$_{z}$ with $z$ = 0.99.\cite{Lei HC}

\begin{figure}[tbp]
\centerline{\includegraphics[scale=0.4]{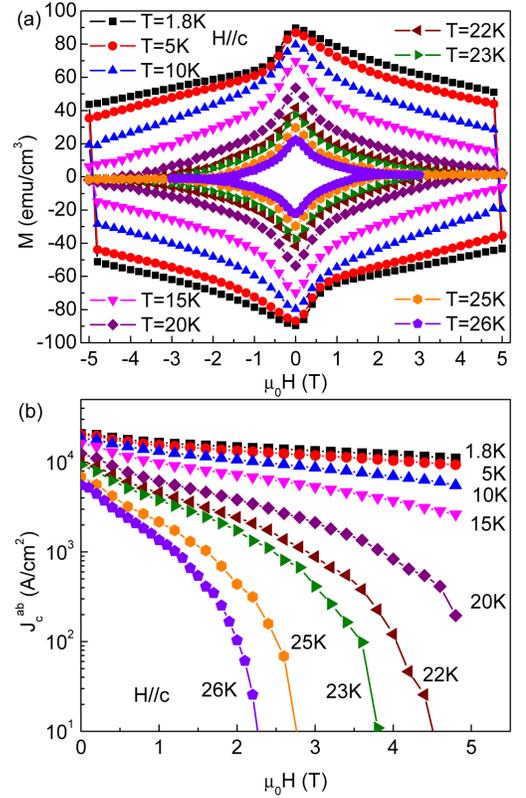}} \vspace*{-0.3cm}
\caption{(a) MHLs of quenched K$_{x}$Fe$_{2-y}$Se$_{2}$ crystal for $%
H\parallel c$. (b) Magnetic field and temperature dependencies of
superconducting critical current densities $J_{c}^{ab}(\protect\mu _{0}H)$
for quenched K$_{x}$Fe$_{2-y}$Se$_{2}$ crystal determined from MHLs.}
\end{figure}

The temperature dependent symmetric curves for all MHLs imply that the bulk
pinning dominates in the crystal at all temperatures. The hysteresis area
decreases with the temperature suggesting gradual decrease of $J_{c}^{ab}$
as the temperature is increased (Fig. 2(b)). The current carrying
performance of quenched crystals is superior at all temperatures and fields
when compared to crystals prepared using the one-step technique.\cite{Gao ZS}

\begin{figure}[tbp]
\centerline{\includegraphics[scale=0.4]{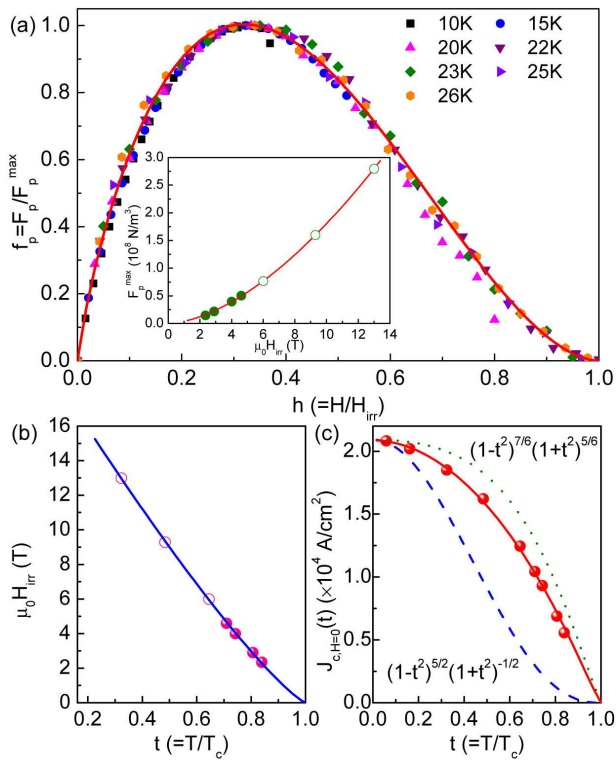}} \vspace*{-0.3cm}
\caption{(a) Reduced field dependence of normalized flux pinning force $%
f_{p} $($h$) at various temperatures. Solid line is the fitting curve using $%
f_{p}=Ah^{p}(1-h)^{q}$. Inset shows $F_{p}^{\max }$ as a function of $%
\protect\mu _{0}H_{irr}$. The fitting result using $F_{p}^{\max }=A(\protect%
\mu _{0}H_{irr})^{\protect\alpha }$ is shown as solid lines. (b) Reduced
temperature dependence of $\protect\mu _{0}H_{irr}$($t$) with the solid line
standing for the fitting result obtained by using the (1-$t$)$^{\protect%
\beta }$ law. (c) Reduced temperature dependence of the $J_{c}$($t$) at zero
field. The dotted, dashed and solid lines show the $J_{c,H=0}^{\protect%
\delta T_{c}}(t)$, $J_{c,H=0}^{\protect\delta l}(t)$ and the fitting result
using $J_{c,H=0}(t)=xJ_{c,H=0}^{\protect\delta T_{c}}(t)+(1-x)J_{c,H=0}^{%
\protect\delta l}(t)$, respectively (see text). The measured and estimated $%
\protect\mu _{0}H_{irr}$ are shown as closed and open circles in inset of
(a) and (b).}
\end{figure}

In order to explain the mechanism of flux pinning in quenched sample, we
studied the temperature and field dependencies of the vortex pinning force $%
F_{p}=\mu _{0}HJ_{c}$. Based on the Dew-Huges model,\cite{Dew-Hughes} if
there is a dominant pinning mechanism then the normalized vortex pinning
forces $f_{p}=F_{p}/F_{p}^{\max }$ from different measurement temperatures
should overlap and a scaling law of the form $f_{p}\propto h^{p}(1-h)^{q}$
will be observed. Here $h$ is the reduced field $h=H/H_{irr}$ and $%
F_{p}^{\max }$ corresponds to the maximum pinning force.\ The
irreversibility field $\mu _{0}H_{irr}$ is the magnetic field where $%
J_{c}^{ab}(T,\mu _{0}H)$ extrapolates to zero. The indices $p$ and $q$
provide the information about the pinning mechanism. As shown in Fig. 3(a),
the normalized curves of $f_{p}(h,T)$ for $T$ $\geqslant $ 22 K present a
temperature independent scaling law. Using the scaling function $%
h^{p}(1-h)^{q}$, we estimate $p$ = 0.86(1) and $q$ = 1.83(2), respectively.
The value of $h_{\max }^{fit}$ ($=p/(p+q)$) $\approx $ 0.32 is consistent
with the peak positions ($h_{max}^{\exp }\approx $ 0.33) of the experimental
curves at different temperatures. Those values are close to expected values
for core normal point-like pinning ($p$ = 1, $q$ = 2, and $h_{\max }^{fit}$
= 0.33).\cite{Dew-Hughes} Moreover, for $T$ $\leqslant $ 20 K, the $H_{irr}$
can be estimated by $F_{p}^{\max }$ location at $h_{max}$ = 0.33. Partial $%
f_{p}(h,T)$ curves measured between 10 and 20 K also exhibit the same
scaling law, suggesting that core normal point-like pinning mechanism is
dominant above 10 K. These point-like pinning center could come from the
random distribution of Fe vacancies after quenching, similar to FeAs-122
type materials.\cite{Yang H}$^{,}$\cite{Sun DL}$^{,}$\cite{Yamamoto} On the
other hand, the $F_{p}^{\max }$ obeys the $F_{p}^{\max }\propto (\mu
_{0}H_{irr})^{\alpha }$ scaling with $\alpha =$ 1.67(1) (inset of Fig.
3(a)), close to the theoretical value ($\alpha $ = 2) for the core normal
point-like pinning.\cite{Dew-Hughes} Moreover, as shown in Fig. 3(b), the
temperature dependence of $\mu _{0}H_{irr}$ can be fitted by using $\mu
_{0}H_{irr}(T)=\mu _{0}H_{irr}(0)(1-t)^{\beta }$ where $t=T/T_{c}$ and we
obtained $\beta $ = 1.21(1), close to the characteristic value of 3D giant
flux creep ($\beta $ = 1.5).\cite{Yeshurun} Similar index has been observed
in overdoped Ba(Fe$_{1-x}$Co$_{x}$)$_{2}$As$_{2}$.\cite{Shen B}

Given the presence of 3D core pinning in quenched K$_{x}$Fe$_{2-y}$Se$_{2}$
single crystals, it is important to distinguish between the case of $\delta
T_{c}$ and $\delta l$ pinnings. For type-II superconductors,\ vortices
interact with pinning centers either via the spatial variations in the $%
T_{c} $ ("$\delta T_{c}$ pinning") or by scattering of charge carriers with
reduced mean free path $l$ near defects ("$\delta l$ pinning").\cite{Blatter}
These two pinning types have different temperature dependence and therefore
result in different relationship between $J_{c}(t)$ and $t=T/T_{c}$ in the
single vortex-pinning regime (low-field and zero-field regions). For $\delta
T_{c}$ pinning, $J_{c,H=0}^{\delta
T_{c}}(t)=J_{c,H=0}(0)(1-t^{2})^{7/6}(1+t^{2})^{5/6}$ while for $\delta l$
pinning, $J_{c,H=0}^{\delta
l}(t)=J_{c,H=0}(0)(1-t^{2})^{5/2}(1+t^{2})^{-1/2} $.\cite{Griessen} As shown
in Fig. 3(c), the $J_{c,H=0}(t)$ is between the two curves corresponding to $%
\delta T_{c}$\ and $\delta l$ pinnings, respectively, but much closer and
similar in shape to the $\delta T_{c}$-pinning curve. Using $%
J_{c,H=0}(t)=xJ_{c,H=0}^{\delta T_{c}}(t)+(1-x)J_{c,H=0}^{\delta l}(t)$, the
experimental data can be fitted very well with $x$ = 0.74(2), suggesting
that both $\delta T_{c}$\ and $\delta l$ pinnings play roles in the quenched
K$_{x}$Fe$_{2-y}$Se$_{2}$ single crystals, but the former mechanism is
dominant. It also implies that the main pinning centers lead to the
distribution of $T_{c}$ in their vicinity or even might be
non-superconducting like Y$_{2}$O$_{3}$ and Y-Cu-O precipitates in YBa$_{2}$%
Cu$_{3}$O$_{7-x}$ thin films.\cite{Ijaduola}

Even though the $J_{c}^{ab}$ of quenched K$_{x}$Fe$_{2-y}$Se$_{2}$ single
crystals is still one or two order(s) smaller than that of other iron
pnictide superconductors,\cite{Yang H}$^{,}$\cite{Sun DL}$^{-}$\cite%
{Prozorov} post-annealing and quenching technique is an effective way to
increase the $J_{c}^{ab}$ of K$_{x}$Fe$_{2-y}$Se$_{2}$.

\section{Conclusion}

In summary, we report giant increase in the $J_{c}^{ab}$ of K$_{x}$Fe$_{2-y}$%
Se$_{2}$ single crystals by post-annealing and quenching technique. We
demonstrate that quenched K$_{x}$Fe$_{2-y}$Se$_{2}$ crystals carry the
highest observed $J_{c}^{ab}$ among FeCh-122 type materials and exhibit good
performance at high field. Detailed analysis of vortex pinning mechanism
points out to the presence of a 3D point-like normal core pinning in
quenched samples. Moreover, the analysis of temperature dependence of $%
J_{c}^{ab}$ at zero field indicates that the $\delta T_{c}$ pinning is
dominant at measured temperature range.

\section{Acknowledgements}

Work at Brookhaven is supported by the U.S. DOE under Contract No.
DE-AC02-98CH10886 and in part by the Center for Emergent Superconductivity,
an Energy Frontier Research Center funded by the U.S. DOE, Office for Basic
Energy Science.

\end{document}